\documentclass[runningheads]{llncs}
\usepackage[T1]{fontenc}
\usepackage{amsmath}
\usepackage{graphicx}
\usepackage[table,xcdraw]{xcolor}
\usepackage[font=small,skip=2pt]{caption}
\usepackage[skip=0pt]{subcaption}
\usepackage{multicol}
\usepackage{multirow}
\usepackage{adjustbox}
\usepackage{hyperref}
\usepackage{color}
\usepackage{makecell}

\begin{document}
\bibliographystyle{splncs04}

\title{Landmark Detection for Medical Images using a General-purpose Segmentation Model}

\titlerunning{Landmark Detection for Medical Images using a General-purpose Segmentation Model}
\author{Ekaterina Stansfield\orcidID{0000-0001-8548-0995}\inst{1} \and
        Jennyfer A. Mitterer\orcidID{0000-0002-5465-3459}\inst{2} \and 
        Abdulrahman Altahhan\orcidID{0000-0003-1133-7744}\inst{3}}
\authorrunning{E. Stansfield, J.A. Mitterer and A. Altahhan}
\institute{
        University of Vienna, Djerassiplatz 1, Vienna, 1030 Austria \\
        \email{katya.stansfield@univie.ac.at} \and
        Orthopaedic Hospital Speising, Speising Str. 109, Vienna, 1130 Austria \\
        \email{JennyferAngel.Mitterer@oss.at} \and
        University of Leeds, Leeds, LS2 9JT United Kingdom \\
        \email{od21ss@leeds.ac.uk, a.altahhan@leeds.ac.uk}}

\maketitle

\begin{abstract}

Radiographic images are a cornerstone of medical diagnostics in orthopaedics, with anatomical landmark detection serving as a crucial intermediate step for information extraction. General-purpose foundational segmentation models, such as SAM (Segment Anything Model), do not support landmark segmentation out of the box and require prompts to function. However, in medical imaging, the prompts for landmarks are highly specific. Since SAM has not been trained to recognize such landmarks, it cannot generate accurate landmark segmentations for diagnostic purposes. Even MedSAM, a medically adapted variant of SAM, has been trained to identify larger anatomical structures—such as organs and their parts—and lacks the fine-grained precision required for orthopaedic pelvic landmarks. To address this limitation, we propose leveraging another general-purpose, non-foundational model: YOLO. YOLO excels in object detection and can provide bounding boxes that serve as input prompts for SAM. While YOLO is efficient at detection, it is significantly outperformed by SAM in segmenting complex structures. In combination, these two models form a reliable pipeline capable of segmenting not only a small pilot set of eight anatomical landmarks but also an expanded set of 72 landmarks and 16 regions with complex outlines—such as the femoral cortical bone and the pelvic inlet. By using YOLO-generated bounding boxes to guide SAM, we trained the hybrid model to accurately segment orthopaedic pelvic radiographs. Our results show that the proposed combination of YOLO and SAM yields excellent performance in detecting anatomical landmarks and intricate outlines in orthopaedic pelvic radiographs. 

\end{abstract}

\section{Introduction and Background}

\begin{figure}[!ht]
    \centering
    \includegraphics[width=0.8\linewidth]{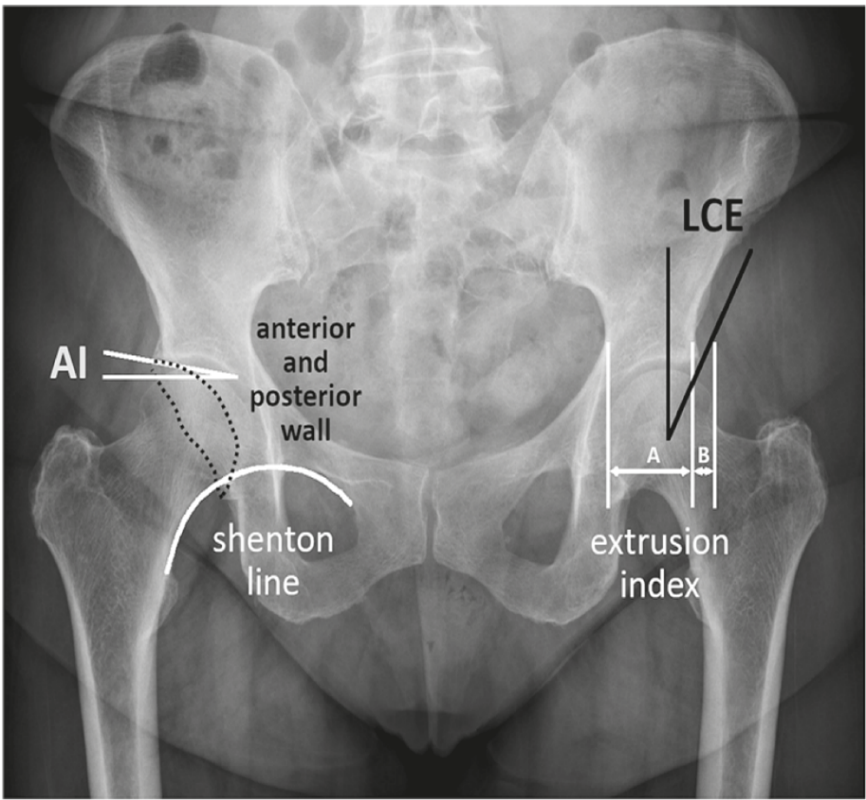}
    \caption{Schematic illustration of radiographic hip parameters on an AP pelvic radiograph: AI – acetabular index, LCE – lateral central edge; A – covered femoral head; B – undercovered femoral head. (Reproduced from Fig.4 by Mitterer et al. (2022)).}
    \label{fig:Fig1}
\end{figure}

Analysis of orthopaedic X-rays relies on a set of angles and ratios that can be directly measured on the radiographs (Fig.~\ref{fig:Fig1}) \cite{tannastHanke2015radiographic}\cite{tannastFritsch2015radiographic}. However, recent trends in the automatization and computerisation of health diagnostics introduced an intermediate step where radiographs are used to identify coordinates of anatomical landmarks (i.e. anatomically meaningful points on images). Relevant angles and ratios are then calculated from the landmark locations \cite{Chai2024}\cite{zhao2024landmark}. Unfortunately, the available commercial software is limited in what data it can produce at a large scale. Simultaneously, time constrains doctors regarding what data they can collect manually.  To continue producing relevant research, they need a flexible and scalable approach that can be re-used with any new set of measurements in emerging projects.

Automated identification of landmarks in medical images has been a thriving direction in AI applications \cite{polizzi2024automatic}. In essence, it is considered a type of image segmentation in which objects are determined by their position relative to image features \cite{ramadan2022convolution}\cite{weingart2023automated}. The majority of recent work implements some form of uNet mechanism \cite{ronneberger2015u} with or without additional attention modules to tokenise images \cite{pei2023learning}\cite{ham2023learning}\cite{schobs_anatomical_2023}\cite{Chong2024-pf}. However, other techniques have also been explored, such as map regression \cite{sanchez2024segmentation} and reinforcement learning \cite{kumar2024anatomical}. In all past cases, researchers conducted supervised training of models from scratch.

Until recently, the primary obstacle to customising AI pipelines for medical image segmentation has been the need for large amounts of labelled training data \cite{ma2023towards}. However, advancements in foundation models \cite{xie2024sam}\cite{liu2024groundingdinomarryingdino}\cite{cheng2024yolo} have opened new possibilities. Computer vision foundation models are trained on millions of images and billions of labels, with the labelling process becoming increasingly automated \cite{ren2024grounding}\cite{ravi2024sam}. These models can be adapted to specific tasks through fine-tuning \cite{ma2023towards} or low-rank adaptation (LoRA) \cite{zhong2024convolution} using significantly smaller training datasets.

For example, Segment Anything Model (SAM) was fine-tuned for segmenting bone structures on MRI images using 300 MRI volumes covering various anatomical regions \cite{gu2024build}. This represents a significantly smaller dataset than what was used to train the original foundation model, demonstrating the efficiency of fine-tuning techniques.

Despite the advances in foundation models for image segmentation, they have not yet been adapted for landmark detection on medical images. This study evaluates the possibility of using foundation models for image segmentation as a basis for landmarking medical X-rays. Unlike previous approaches, which typically rely on task-specific models trained from scratch or with limited transfer learning, this research explores the novel application of large-scale pre-trained vision models for precise anatomical landmark detection. This could reduce the need for extensive labelled datasets while improving accuracy and generalizability.

\section{Design and Methods}

\subsection{Problem outline}
The dataset available for the present project comprised 100 anonymised frontal radiographs of the human pelvis, provided by Speising Orthopaedic Hospital, Vienna. These radiographs were annotated with 72 individual landmarks and a further set of landmarks around 18 patches and outlines. The radiographs were in the original DICOM format, while annotations were supplied separately as JSON files. 
Since the size of the sample was small, the model used in this study needed to possess prior domain knowledge or be easily fine-tuned to the new domain. The detected landmarks had to exhibit high accuracy and precision, as they would be used in clinical decision-making. Each landmark had to belong to a distinct class. Furthermore, the chosen pipeline needed to be scalable to accommodate a combination of landmarks, outlines, and solid masks, even when these elements overlapped. Lastly, the process had to require minimal specialist knowledge of artificial intelligence engineering, ensuring that it could be modified and maintained by operators with limited training in a hospital setting.

\subsection{Models and Strategy Choice}
The ‘vanilla’ versions of open-source models frequently require high-end resources for finetuning. For example, MedSam was finetuned with eight A100 Nvidia cards and 80GB memory each \cite{ma2024segment}\cite{ma2023towards}, which are not available to an average user. Alternatives exist in the form of open-source application programming interfaces (APIs). Huggingface API (https://huggingface.co/facebook/sam-vit-base, accessed January 2025) offers adapted models requiring fewer resources and over 1,000 pre-trained models, while Ultralitics provides free, user-friendly You Only Look Once (YOLO) API (https://www.ultralytics.com/, accessed November 2024).
Upon consideration, Ultralithics YOLO 11 was chosen as the main candidate for locating landmarks. The latest version of this pipeline includes a range of neural networks and attention modules, offers enhanced feature extraction, is optimised for efficiency and speed, and provides greater accuracy with fewer parameters (https://docs.ultralytics.com/models/yolo11/, accessed November 2024). Importantly, it does not require large and expensive resources for fine-tuning and can be trained on an average laptop with one NVIDIA RTX 3050 video card. It was possible to fine-tune this model for a completely unfamiliar domain.
The second model used here was the Huggingface version of the Segment Anything Model (SAM) (https://huggingface.co/facebook/sam-vit-base, https://huggingface.co/flaviagiammarino/medsam-vit-base, accessed November 2024) with the weights published by the creators of MedSam \cite{ma2023towards}. These weights allowed SAM to recognise features on medical images so that no further domain fine-tuning of the encoder was required. It was, therefore, possible to fine-tune only SAM’s decoder, reducing the amount of the parameters to train and using fewer resources and less compute time.
Unlike YOLO, SAM cannot classify detected masks. However, thanks to the vast amount of data used for its pre-training, it has a superior capacity for accurate and precise image segmentation. It can still be used if YOLO provides bounding boxes as prompts to locate the desired mask for segmentation.

\subsection{Data choice and sample split}
The experiment was set to run in two stages. To match previous publications and to reduce the training time, only eight out of 72 available single landmarks, and no patches or outlines were used at the first stage. At the second stage, the problem was scaled up to the whole set of available data.
The available dataset was programmatically transformed to create NumPy arrays for each of the 72 landmarks and 18 outlines and patches. These data were subsequently used to develop model-specific labels, such as outlines and bounding boxes for YOLO and tensors for SAM. Eighty radiographs were used for training, five for validation, and 15 were kept back as unseen test samples.  

\subsection{Results evaluation}
Accuracy and precision are two critical parameters for the present study. Accuracy has been inferred from the Intersection over Union statistics (IoU) provided by the YOLO pipeline or using the original script to evaluate the accuracy of the SAM-predicted labels. Precision has been interpreted as a point-to-point distance between the prediction and the ground truth. Precision, therefore, was only possible to calculate for landmark data. The acceptable level of the predicted landmark error distance from the ground truth was set at 3 mm. Patches and outlines, as introduced in the final part of the study, were only assessed via the IoU.

\section{Results}

\subsection{Ground-base uNet, eight landmarks}
\begin{figure}[!ht]
    \centering
    \includegraphics[width=0.8\linewidth]{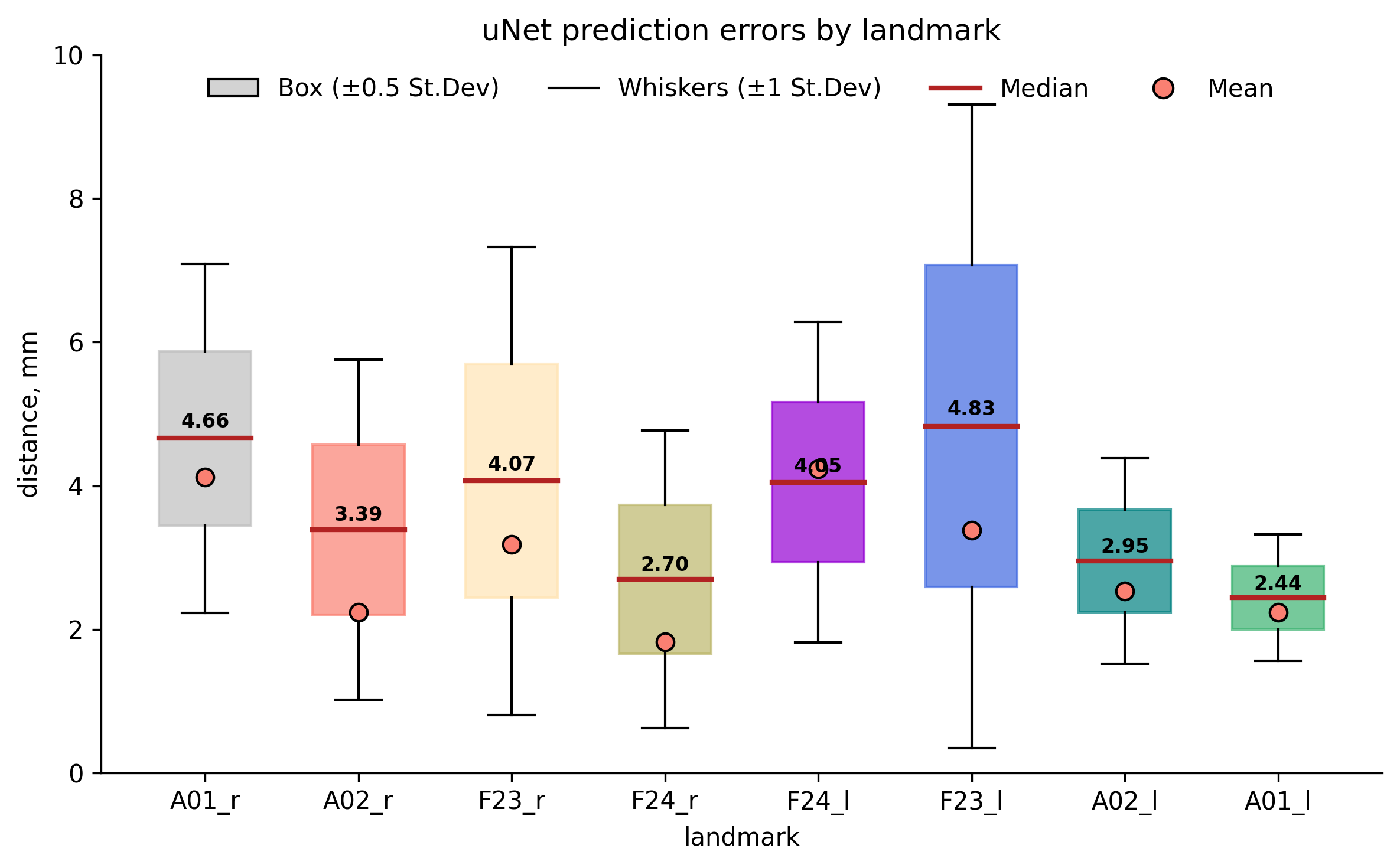}
    \caption{uNet prediction errors (mm) for the test individuals. A01 acetabulum medial sourcil, A02 acetabular lateral sourcil, F23 centre of hip rotation, F24 femoral neck midpoint: \_r and \_l signifies right and left side accordingly.}
    \label{fig:Fig2}
\end{figure}

The architecture used here belongs to the original u-Net \cite{ronneberger2015u}, adapted from Schobs  dissertation and available at https://github.com/Schobs/MediMarker (accessed April 2024) \cite{schobs_anatomical_2023}. It included two blocks per resolution layer, each of 3x3 convolution, Instance Normalisation and Leaky ReLU (slope -0.01). Downsampling was achieved through strided convolutions and upsampling through transposed convolutions. Input image resolution was 512, and the initial number of feature maps was 32, doubling with each downsample. The number of resolution layers was configured automatically by adding encoder steps. The loss was measured with the help of the Mean Standard Error between the Gaussian target heatmap and the predicted heatmap. Standard deviation was used as a hyperparameter. Deep supervision was achieved by injecting losses at every layer. Precision was evaluated using point-to-point Eucledian distance (Fig.~\ref{fig:Fig2}).
The sample was augmented by rotation, translation, shearing and scaling but not reflection, as the latter introduced significant errors in the bilateral set of predicted labels. Training occurred over 500 epochs, using 8 CPU cores and eight shared T1 GPUs, 20GB memory for 26 hours of a computer cluster time.

The resultant mean and median error values (Tab. 1) were comparable with previous publication by Pei et al.\cite{pei2023learning}.

\subsection{YOLO segmentation, eight landmarks}
\begin{figure}[!ht]
    \centering
    \includegraphics[width=0.8\linewidth]{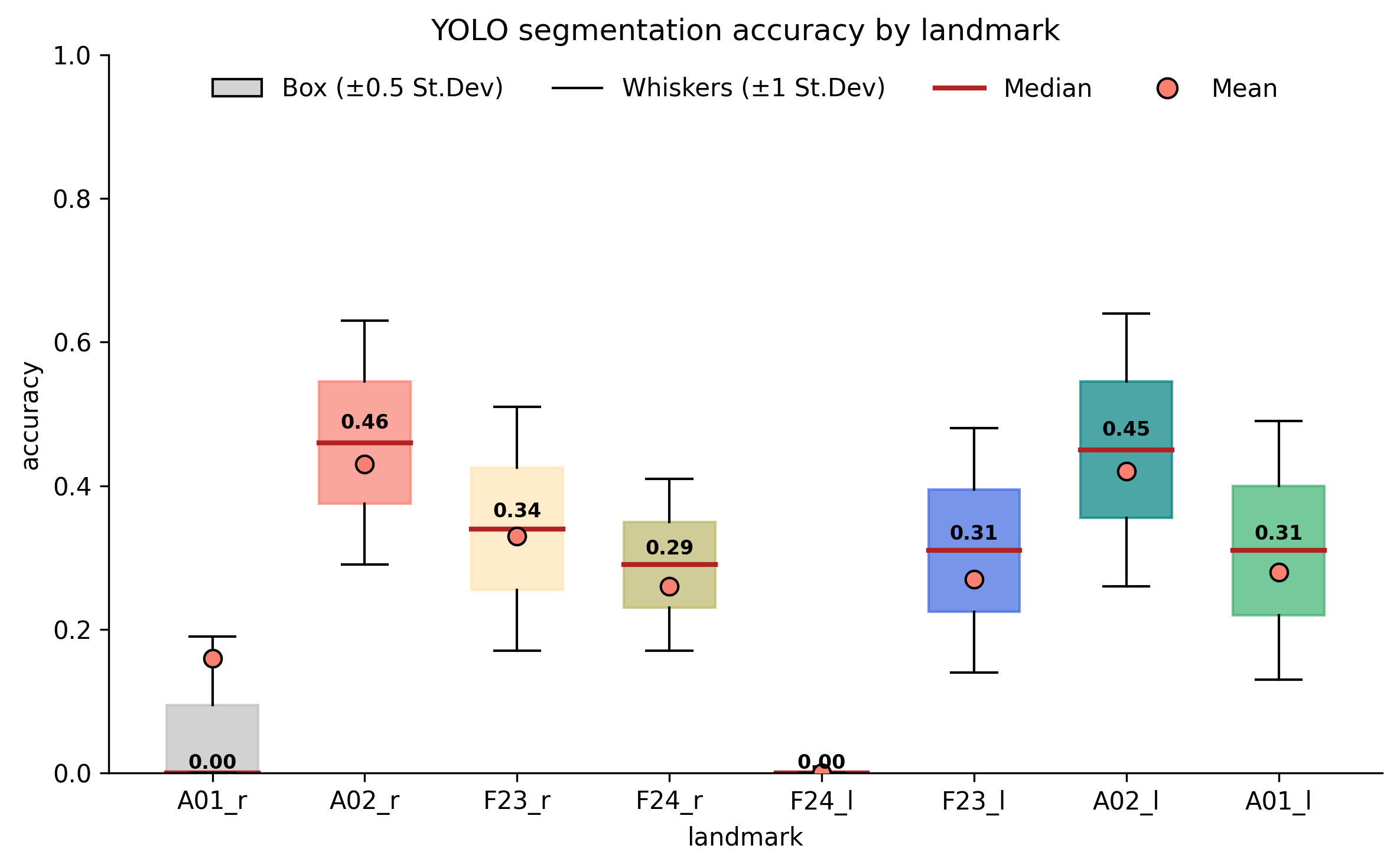}
    \caption{Accuracy of prediction for test individuals.}
    \label{fig:Fig3}
\end{figure}

YOLO11-s is a model with 10.1M trainable parameters. Training was done over 300 epochs with sample augmentation by varying brightness, contrast, translation, scaling and angle variation without reflection. Training took 30 minutes on a PC with one NVIDIA RTX 3050 GPU. The accuracy of results, however, was low (Fig.~\ref{fig:Fig3}). Therefore, YOLO segmentation was not used in further experiments.

\subsection{YOLO detection, eight landmarks}
YOLO11-n is a model with 6.5M trainable parameters. Training was done over 300 epochs with sample augmentation by variation of brightness, contrast, translation, scaling and angle variation without reflection. Training took 30 minutes on a PC with one NVIDIA RTX 3050 GPU.

\begin{figure}[!ht]
    \centering
    \includegraphics[width=0.8\linewidth]{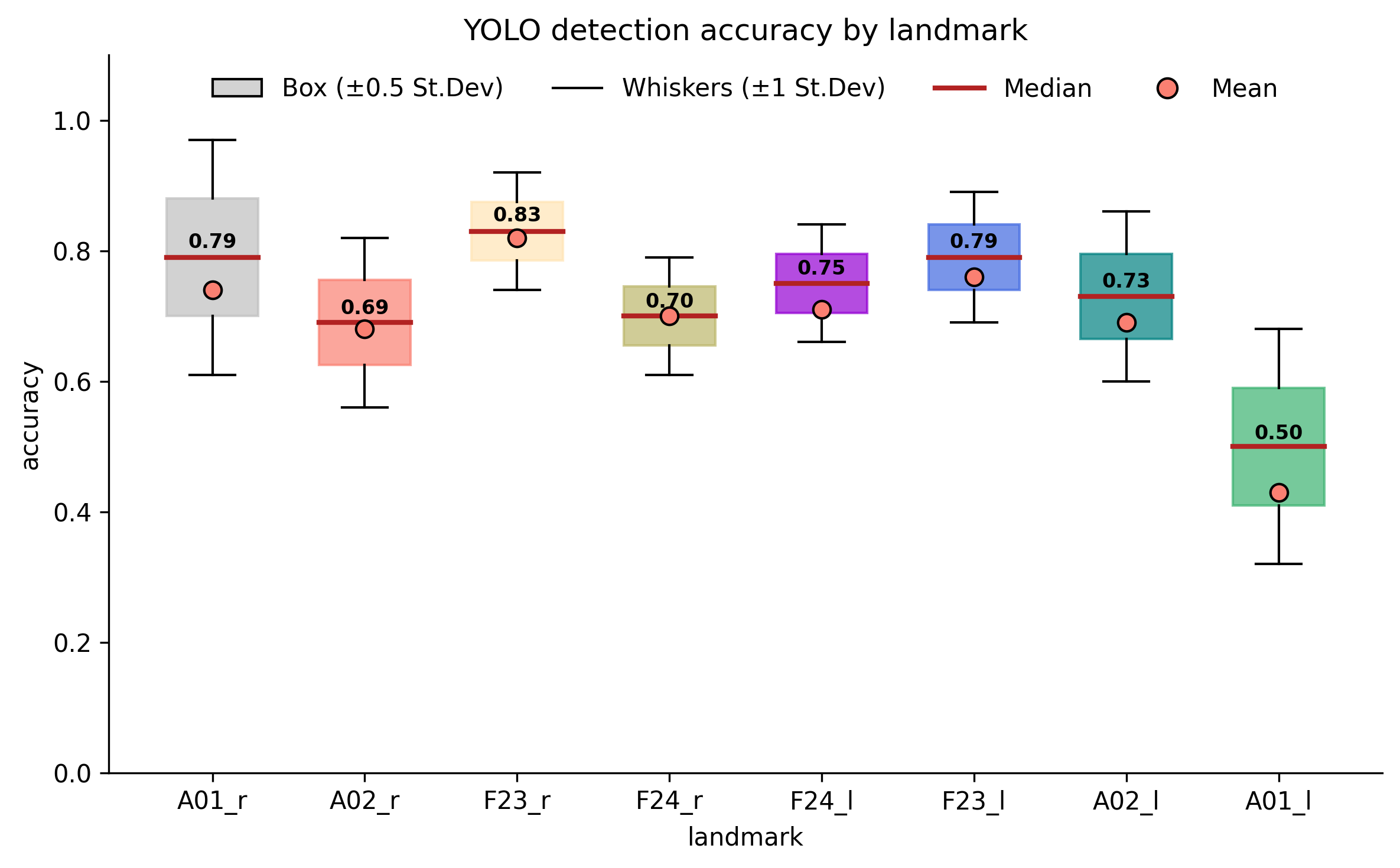}
    \caption{Accuracy of predictions for test individuals.}
    \label{fig:Fig4}
\end{figure}

\begin{figure}[!ht]
    \centering
    \includegraphics[width=0.8\linewidth]{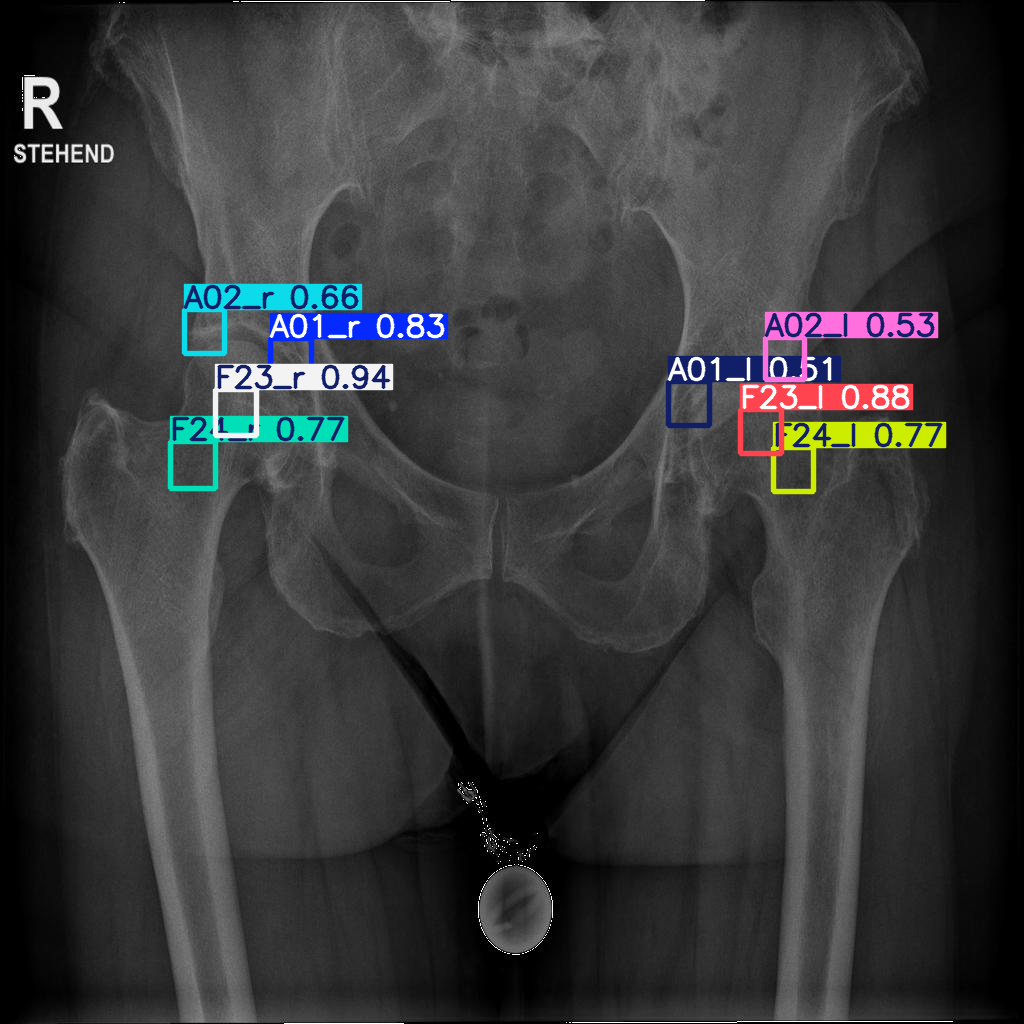}
    \caption{Successful identification of all landmarks in a test image.}
    \label{fig:Fig5}
\end{figure}

The mean and median accuracies were significantly better than that of the segmentation algorithm (Fig.~\ref{fig:Fig4}, Fig.~\ref{fig:Fig5}). The lowest median accuracy across all landmarks was 0.5.

\begin{figure}[!ht]
    \centering
    \includegraphics[width=0.8\linewidth]{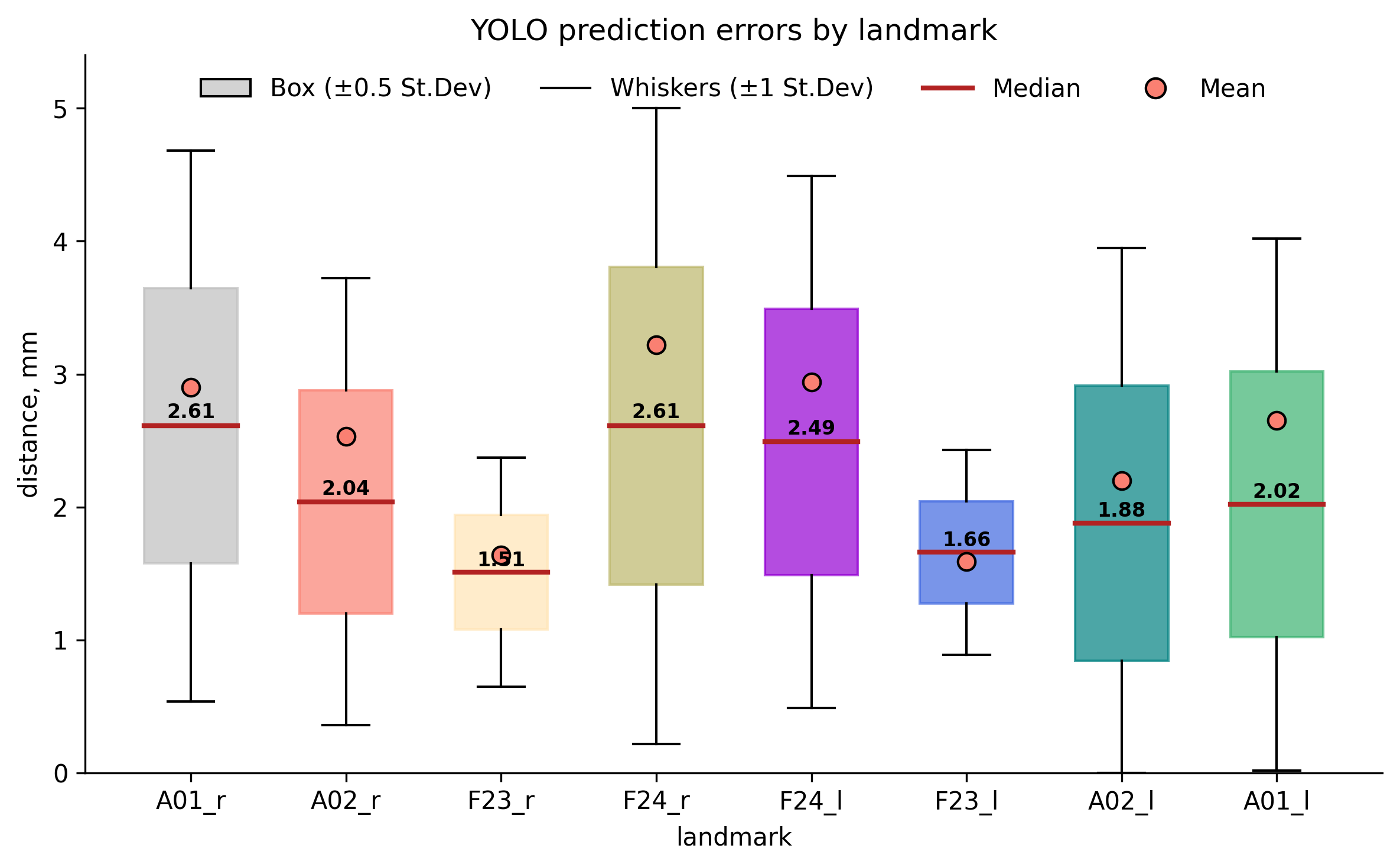}
    \caption{Prediction errors (mm) for the test individuals.}
    \label{fig:Fig6}
\end{figure}

Predicted bounding boxes were used to find the coordinates of their centres. Interpreted as landmark locations, they were transformed to the original image dimensions. Error point-to-point distances to the ground truth were calculated in mm (Fig.~\ref{fig:Fig6}). The median errors were within the acceptable range for medical image analysis (3 mm).

\subsection{Scaled problem on 72 landmarks, 18 patches and outlines}
\begin{figure}[!ht]
    \centering
    \includegraphics[width=0.8\linewidth]{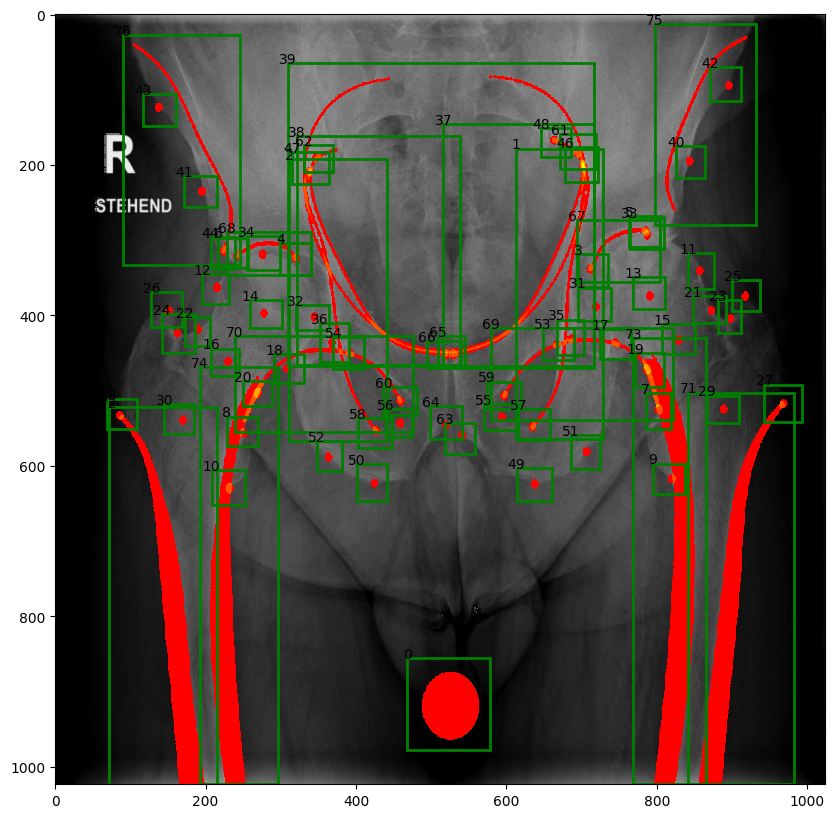}
    \caption{Ground truth data, bounding boxes, landmarks, outlines and patches.}
    \label{fig:Fig7}
\end{figure}

For the scaled problem, the landmarks were defined as circles of 2 mm radius. An outline was represented by an orthopaedic line, created by the superimposition of 3D features on a radiograph. Patches filled in cortical bone of the left and right femora and the calibration ball in the middle. Some of the features overlap in part (Fig.~\ref{fig:Fig7}).

\subsection{YOLO detection, the scaled task}
YOLO11-n model was used again in the scaled task. As in the case with eight landmarks, training was carried out for 300 epochs with sample augmentation by variation of brightness, contrast, translation, scaling and angle variation without reflection. Time of training was 30 min.
The mean accuracy of landmark determination was lower in the expanded dataset (Tab. 1). Five out of 72 landmarks, i.e. 7\%, were not identified. Most of these points were closely positioned on the radiograph and, hence, difficult to differentiate. The mean accuracy of patches, on the other hand, was high despite two out of 18 outlines, i.e. 11\%, not located at all.

\begin{table}
    \centering
    \begin{tabular}{|>{\centering\arraybackslash}p{0.15\linewidth}|>{\centering\arraybackslash}p{0.2\linewidth}|>{\centering\arraybackslash}p{0.2\linewidth}|>{\centering\arraybackslash}p{0.2\linewidth}|} \hline 
     & LANDMARKS on femora & LANDMARKS on pelvis & PATCHES AND OUTLINES \\ \hline 
      median   & 0.46 & 0.49 & 0.83\\ \hline 
       mean  & 0.46 & 0.46 & 0.83\\ \hline 
       st.dev  & 0.13 & 0.19 & 0.05\\ \hline
    \end{tabular}
    \caption{Accuracy of results for test individuals.}
    \label{tab:Tab1}
\end{table}

\subsection{Training SAM to use YOLO-predicted boxes for reconstructing landmarks, outlines and patches}

\begin{figure}[!ht]
    \centering
    \includegraphics[width=0.8\linewidth]{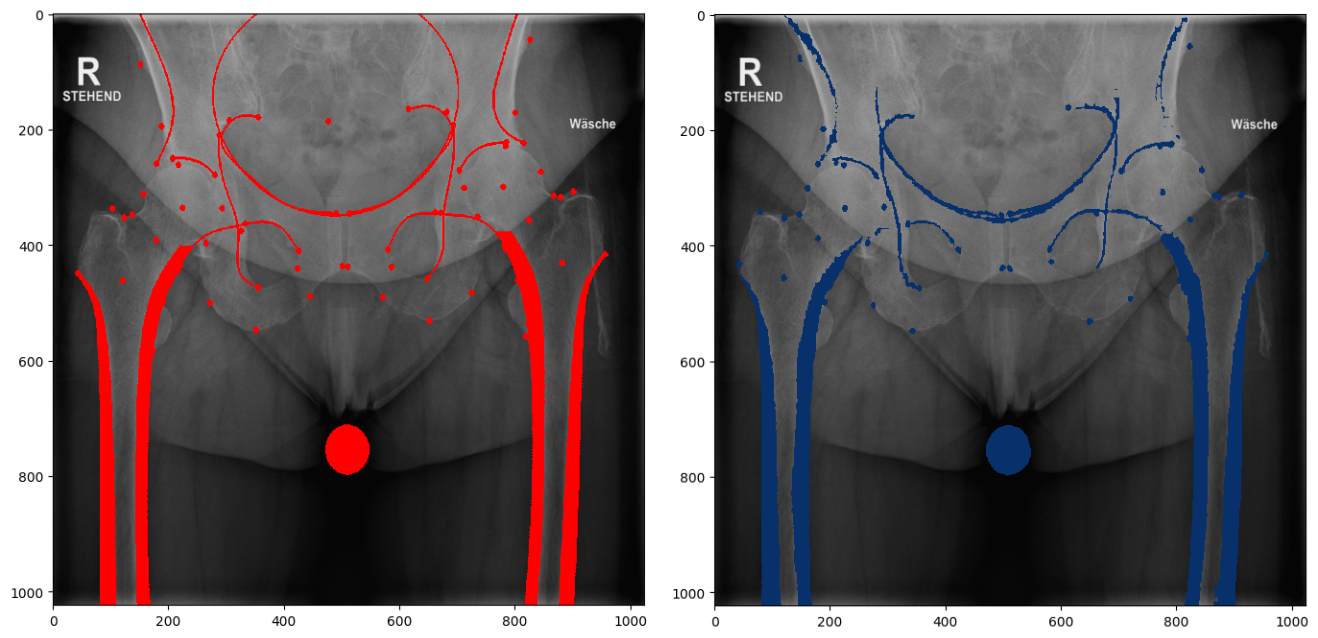}
    \caption{Identification of landmarks and patches, ground truth in red, prediction in blue.}
    \label{fig:Fig8}
\end{figure}

Training resources for SAM involved a computer cluster with a dedicated L40 NVIDIA, 8 CPU cores per task, and 200 GB memory for 36 hours. Figure~\ref{fig:Fig4} shows a typical successful result of identification. Figure~\ref{fig:Fig8} givens an example of a ground truth and a predicted segmentation.
Results for 15 test cases, landmarks: 
\begin{itemize}
    \item Median error (distance in mm) for identified landmarks: 1.66 mm
    \item Mean error for identified landmarks: 2.30 mm
    \item Standard deviation of the error: 1.77 mm
\end{itemize}
Results for 15 test cases, patches and outlines: 
\begin{itemize}
    \item Median IoU for identified items: 0.74
    \item Mean IoU for identified items: 0.77
    \item Standard deviation: 0.08
\end{itemize}

\section{Discussion}

This study tested the feasibility of using computer vision general-purpose models to identify and segment landmarks in medical images. The models used here were fine-tuned to recognise features on radiographs and output labels, as in YOLO, or just output the required labels, as in SAM. The exact choice of models was partly by the possibility of fine-tuning without deploying large-volume, expensive computing resources. This was readily possible with YOLO, while SAM required 36 hours of a computer cluster time with multiple CPUs and one 40GB GPU involved.

The accuracy and precision of labels, predicted by trained pipelines on an unseen test sample, were set as the primary benchmark for the training success. In this respect, the results can be compared with previous publications on automatic landmark identification. For example, Pei et al. \cite{pei2023learning} reported results for pelvic radiographs. These authors collected eight landmarks from 524 images in an open-source project supported by the Foundation for the National Institute of Health (https://www.niams). nih.gov/grants-funding/funded-research/osteoarthritis- initiative). YOLO detection in the current study (Fig.~\ref{fig:Fig6}) outperformed both, the uNet that was trained on the same data as YOLO here (Fig.~\ref{fig:Fig2}), and all the models tested by Pei et al. \cite{pei2023learning} (Tab. 2). 

Table 2.

\begin{table}
    \centering
    \begin{tabular}{|>{\centering\arraybackslash}p{0.15\linewidth}|>{\centering\arraybackslash}p{0.1\linewidth}|>{\centering\arraybackslash}p{0.1\linewidth}|>{\centering\arraybackslash}p{0.1\linewidth}|>{\centering\arraybackslash}p{0.1\linewidth}|>{\centering\arraybackslash}p{0.1\linewidth}|>{\centering\arraybackslash}p{0.1\linewidth}|>{\centering\arraybackslash}p{0.1\linewidth}|>{\centering\arraybackslash}p{0.1\linewidth}|} \hline 
         & \makecell{Left \\ Tear} & \makecell{Right \\ Tear}
         & \makecell{Left \\ UAR**}
         & \makecell{Right \\ UAR}
         & \makecell{Left \\ IPSS}
         & \makecell{Right \\ IPSS}
         & \makecell{Left \\ FHC}
         & \makecell{Right \\ FHC}
\\ \hline 
      HR-Net*   & 3.3753 & 3.8895 & 2.7483 & 2.7324 & 3.9612 & 3.9317 & 3.0886 & 3.3856\\ \hline 
       CE-Net  & 3.0493 & 3.1254 & 3.0806 & 2.4243 & 3.8111 & 3.3232 & 3.0433 & 3.3254\\ \hline 
      uNet with attention   & 2.911 & 3.2465 & 2.9164 & 2.3435 & 3.432 & 3.44 & 2.4504 & 2.9378\\ \hline
    \end{tabular}
    \caption{verage point-to-point distances from the ground truth (from Tab. 2. Pei et al.\cite{pei2023learning}). *HR-Net - high-resolution representation neural network; CE-Net - a uNet with a replaced encoder module with ResNet residual structure; uNet with attention - originally proposed by Pei et al. \cite{pei2023learning}
**UAR - upper acetabulum rim, IPSS - innermost point of subchondral sclerosis, FHC - femur head centre.
}
    \label{tab:Tab2}
\end{table}

YOLO detection worked well for localising landmarks, provided that the obtained bounding box was used to calculate its centre, i.e., the predicted landmark. YOLO segmentation, on the other hand, did not provide adequate results on the original eight landmarks and, therefore, would not have been a sensible choice for scaling the task. Scaling was achieved with the help of a combination of two foundation models. YOLO was used to detect the location of desired features, to provide bounding boxes, and to class identification. The resultant boxes were presented to SAM for drawing the pixel location of the desired feature, whether it was a landmark, an outline or a patch. In production, the two pipelines can be seamlessly combined, so the user is only presented with the suggested prediction for feature locations.

The experiment of scaling to the larger dataset returned mixed results. The average accuracy was relatively low for the landmarks, as some were consistently absent from the predicted features despite their presence in the ground truth (Tab. 5). However, the precision (mean error) of identified landmarks was still only 2.3 mm, within the accepted level of a 3 mm point-to-point distance. The total rate of identified features was 93\% for landmarks and 89\% for patches and outline.  

The results of the current pipeline for a scaled-up set of labels can be enhanced by increasing the size of the training dataset. The 100 radiographs provided for this study were insufficient for YOLO to reliably differentiate between some of the closely positioned points, causing several to be completely missed. Fortunately, the developed pipeline allows for a gradual accumulation of annotated data. AI-generated labels can be reviewed by a medical practitioner through a user interface, enabling error correction and filling in missing data. The human-curated results can then be used for iterative fine-tuning of the models, following the approach of Mahmoodi et al.\cite{mahmoodi2023study}, where newly accumulated data is continuously added to the training pool for subsequent finetuning. This setup will be easily adaptable to more advanced foundation models as they become available, requiring minimal programming effort from the user. Furthermore, the relatively small initial dataset needed for fine-tuning creates opportunities for entirely different projects.

\section{Conclusions}

The YOLO 11 model performed exceptionally well in the detection task but not in segmentation. When YOLO detection was augmented by SAM's outstanding ability to segment desired objects in medical images, it produces a workable set of classified labels that can be recognised and used by a human user.

\bibliography{main}

@Article{Chai2024,
author={Chai, Yuan
and Boudali, A. Mounir
and Maes, Vincent
and Walter, William L.},
title={Clinical benchmark dataset for AI accuracy analysis: quantifying radiographic annotation of pelvic tilt},
journal={Scientific Data},
year={2024},
month={Oct},
day={22},
volume={11},
number={1},
pages={1162},
issn={2052-4463},
doi={10.1038/s41597-024-04003-7},
url={https://doi.org/10.1038/s41597-024-04003-7}
}

@misc{cheng2024yolo,
      title={YOLO-World: Real-Time Open-Vocabulary Object Detection}, 
      author={Tianheng Cheng and Lin Song and Yixiao Ge and Wenyu Liu and Xinggang Wang and Ying Shan},
      year={2024},
      eprint={2401.17270},
      archivePrefix={arXiv},
      primaryClass={cs.CV},
      url={https://arxiv.org/abs/2401.17270}, 
}

@ARTICLE{Chong2024-pf,
  title    = "Automated anatomical landmark detection on {3D} facial images
              using {U-NET-based} deep learning algorithm",
  author   = "Chong, Yuming and Du, Fengzhou and Ma, Xuda and An, Yicheng and
              Huang, Qi and Long, Xiao and Huang, Jiuzuo and Li, Zhijin and Yu,
              Nanze and Wang, Xiaojun",
  journal  = "Quant Imaging Med Surg",
  volume   =  14,
  number   =  3,
  pages    = "2466--2474",
  month    =  mar,
  year     =  2024,
  address  = "China",
  language = "en"
}

@article{gu2024build,
  title={How to build the best medical image segmentation algorithm using foundation models: a comprehensive empirical study with segment anything model},
  author={Gu, Hanxue and Dong, Haoyu and Yang, Jichen and Mazurowski, Maciej A},
  journal={arXiv preprint arXiv:2404.09957},
  year={2024}
}

@article{ham2023learning,
  title={Learning Spatial Configuration Feature for Landmark Localization in Hand X-rays},
  author={Ham, Gyu-Sung and Oh, Kanghan},
  journal={Electronics},
  volume={12},
  number={19},
  pages={4038},
  year={2023},
  publisher={MDPI}
}

@article{kumar2024anatomical,
  title={Anatomical Landmark Detection in 3d MRI Scan using Deep Neuro-Dynamic Programming},
  author={Kumar, Yogesh and Kumar, Pankaj},
  journal={Procedia Computer Science},
  volume={235},
  pages={1713--1721},
  year={2024},
  publisher={Elsevier}
}

@misc{liu2024groundingdinomarryingdino,
      title={Grounding DINO: Marrying DINO with Grounded Pre-Training for Open-Set Object Detection}, 
      author={Shilong Liu and Zhaoyang Zeng and Tianhe Ren and Feng Li and Hao Zhang and Jie Yang and Qing Jiang and Chunyuan Li and Jianwei Yang and Hang Su and Jun Zhu and Lei Zhang},
      year={2024},
      eprint={2303.05499},
      archivePrefix={arXiv},
      primaryClass={cs.CV},
      url={https://arxiv.org/abs/2303.05499}, 
}

@inproceedings{mahmoodi2023study,
  title={A Study on Reducing Big Data Image Annotation Burden Through Iterative Expert-In-The-Loop Strategy},
  author={Mahmoodi, Evanjelin and Xue, Zhiyun and Rajaraman, Sivaramakrishnan and Antani, Sameer},
  booktitle={2023 IEEE International Conference on Bioinformatics and Biomedicine (BIBM)},
  pages={3097--3102},
  year={2023},
  organization={IEEE}
}

@article{ma2024segment,
  title={Segment anything in medical images},
  author={Ma, Jun and He, Yuting and Li, Feifei and Han, Lin and You, Chenyu and Wang, Bo},
  journal={Nature Communications},
  volume={15},
  number={1},
  pages={654},
  year={2024},
  publisher={Nature Publishing Group UK London}
}

@article{ma2023towards,
  title={Towards foundation models of biological image segmentation},
  author={Ma, Jun and Wang, Bo},
  journal={Nature Methods},
  volume={20},
  number={7},
  pages={953--955},
  year={2023},
  publisher={Nature Publishing Group US New York}
}

@article{pei2023learning,
  title={Learning-based landmark detection in pelvis x-rays with attention mechanism: data from the osteoarthritis initiative},
  author={Pei, Yun and Mu, Lin and Xu, Chuanxin and Li, Qiang and Sen, Gan and Sun, Bin and Li, Xiuying and Li, Xueyan},
  journal={Biomedical Physics \& Engineering Express},
  volume={9},
  number={2},
  pages={025001},
  year={2023},
  publisher={IOP Publishing}
}

@article{polizzi2024automatic,
  title={Automatic cephalometric landmark identification with artificial intelligence: An umbrella review of systematic reviews},
  author={Polizzi, Alessandro and Leonardi, Rosalia},
  journal={Journal of Dentistry},
  pages={105056},
  year={2024},
  publisher={Elsevier}
}

@article{ramadan2022convolution,
  title={Convolution neural network based automatic localization of landmarks on lateral x-ray images},
  author={Ramadan, Rabie A and Khedr, Ahmed Y and Yadav, Kusum and Alreshidi, Eissa Jaber and Sharif, Md Haidar and Azar, Ahmad Taher and Kamberaj, Hiqmet},
  journal={Multimedia Tools and Applications},
  volume={81},
  number={26},
  pages={37403--37415},
  year={2022},
  publisher={Springer}
}

@article{ravi2024sam,
  title={Sam 2: Segment anything in images and videos},
  author={Ravi, Nikhila and Gabeur, Valentin and Hu, Yuan-Ting and Hu, Ronghang and Ryali, Chaitanya and Ma, Tengyu and Khedr, Haitham and R{\"a}dle, Roman and Rolland, Chloe and Gustafson, Laura and others},
  journal={arXiv preprint arXiv:2408.00714},
  year={2024}
}

@article{ren2024grounding,
  title={Grounding dino 1.5: Advance the" edge" of open-set object detection},
  author={Ren, Tianhe and Jiang, Qing and Liu, Shilong and Zeng, Zhaoyang and Liu, Wenlong and Gao, Han and Huang, Hongjie and Ma, Zhengyu and Jiang, Xiaoke and Chen, Yihao and others},
  journal={arXiv preprint arXiv:2405.10300},
  year={2024}
}

@inproceedings{ronneberger2015u,
  title={U-net: Convolutional networks for biomedical image segmentation},
  author={Ronneberger, Olaf and Fischer, Philipp and Brox, Thomas},
  booktitle={Medical image computing and computer-assisted intervention--MICCAI 2015: 18th international conference, Munich, Germany, October 5-9, 2015, proceedings, part III 18},
  pages={234--241},
  year={2015},
  organization={Springer}
}

@article{sanchez2024segmentation,
  title={Segmentation-guided coordinate regression for robust landmark detection on X-rays: application to automated assessment of lower limb alignment},
  author={Sanchez, Sebastian Amador and Van Overschelde, Philippe and Vandemeulebroucke, Jef},
  journal={IEEE Access},
  year={2024},
  publisher={IEEE}
}

@phdthesis{schobs_anatomical_2023,
	type = {phd},
	title = {Anatomical {Landmark} {Localisation} and {Uncertainty} {Estimation}},
	copyright = {cc\_by\_nc\_nd\_4},
	url = {https://etheses.whiterose.ac.uk/id/eprint/34822/},
	language = {en},
	urldate = {2025-04-30},
	school = {University of Sheffield},
	author = {Schobs, Lawrence},
	month = sep,
	year = {2023},
}

@article{tannastHanke2015radiographic,
  title={What are the radiographic reference values for acetabular under-and overcoverage?},
  author={Tannast, Moritz and Hanke, Markus S and Zheng, Guoyan and Steppacher, Simon D and Siebenrock, Klaus A},
  journal={Clinical Orthopaedics and Related Research{\textregistered}},
  volume={473},
  pages={1234--1246},
  year={2015},
  publisher={Springer}
}

@article{tannastFritsch2015radiographic,
  title={Which radiographic hip parameters do not have to be corrected for pelvic rotation and tilt?},
  author={Tannast, Moritz and Fritsch, Stefan and Zheng, Guoyan and Siebenrock, Klaus A and Steppacher, Simon D},
  journal={Clinical Orthopaedics and Related Research{\textregistered}},
  volume={473},
  pages={1255--1266},
  year={2015},
  publisher={Springer}
}

@inproceedings{zhao2024landmark,
  title={A landmark-based approach for instability prediction in distal radius fractures},
  author={Zhao, Yang and Liao, Zhibin and Liu, Yunxiang and Oude Nijhuis, Koen and Barvelink, Britt and Prijs, Jasper and Colaris, Joost and Wijffels, Mathieu and Reijman, Max and Zhang, Zeyu and others},
  booktitle={2024 IEEE International Symposium on Biomedical Imaging (ISBI)},
  pages={1--5},
  year={2024},
  organization={IEEE}
}

@article{weingart2023automated,
  title={Automated detection of cephalometric landmarks using deep neural patchworks},
  author={Weingart, Julia Vera and Schlager, Stefan and Metzger, Marc Christian and Brandenburg, Leonard Simon and Hein, Anna and Schmelzeisen, Rainer and Bamberg, Fabian and Kim, Suam and Kellner, Elias and Reisert, Marco and others},
  journal={Dentomaxillofacial Radiology},
  volume={52},
  number={6},
  pages={20230059},
  year={2023},
  publisher={Oxford University Press}
}

@inproceedings{xie2024sam,
  title={Sam fewshot finetuning for anatomical segmentation in medical images},
  author={Xie, Weiyi and Willems, Nathalie and Patil, Shubham and Li, Yang and Kumar, Mayank},
  booktitle={Proceedings of the IEEE/CVF winter conference on applications of computer vision},
  pages={3253--3261},
  year={2024}
}

@article{zhong2024convolution,
  title={Convolution meets lora: Parameter efficient finetuning for segment anything model},
  author={Zhong, Zihan and Tang, Zhiqiang and He, Tong and Fang, Haoyang and Yuan, Chun},
  journal={arXiv preprint arXiv:2401.17868},
  year={2024}
}
\end{document}